# A Test of HTS Power Cable in a Sweeping Magnetic Field

Henryk Piekarz, Steven Hays, Jamie Blowers and Vladimir Shiltsev

*Abstract*—Short sample HTS power cable composed of multiple 344C-2G strands and designed to energize a fast-cycling dipole magnet was exposed to a sweeping magnetic field in the (2-20) T/s ramping rate. The B-field orientation toward the HTS strands wide surface was varied from $0^0$ to $10^0$, in steps of $1^0$. The test arrangement allowed measurement of the combined hysteresis and eddy current power losses. For the validity of these measurements, the power losses of a short sample cable composed of multiple LTS wire strands were also performed to compare with the known data. The test arrangement of the power cable is described, and the test results are compared with the projections for the eddy and hysteresis power losses using the fine details of the test cable structures.

## I. INTRODUCTION

THIS paper is the followed-up of an earlier study [1] aimed at using the HTS strand of a tape form for the construction of transmission-line power cable energizing the fast-cycling accelerator magnet. The power losses induced in the cable by the sweeping magnetic field in the magnet gap of a window-frame shape may be much reduced by orienting the wide side of the tape to be as parallel as possible to the magnetic field in the magnet cable space. The field simulations have shown that a $2^0$ average divergence of the field from the parallel plane of the HTS stack is achievable.

The maximum allowable critical temperature also plays a very significant role as an excessive temperature rise may lead to the instabilities and possibly a quench. The maximum critical temperature of the LTS conductors is 9.2 K for the NbTi and 18.3 K for the $Nb_3Sn$, while for the HTS types, such as e.g. YBCO, it extends up to about 90 K. For the cable operating temperature the allowable temperature margin depends on the ratio of the transport current to the critical current one ($I_t / I_c$). Assuming that all superconductor types are operating at a temperature $T_t = 4.2$ K with the $I_t / I_c = 0.5$ the projected allowable temperature margins are: 1 K for NbTi, 5 K for $Nb_3Sn$ and 28 K for YBCO (e.g. 348C-2G [2]), as illustrated in Fig.1. In order to secure stability of operations the heat generated in the cable must be removed with a very high efficiency.

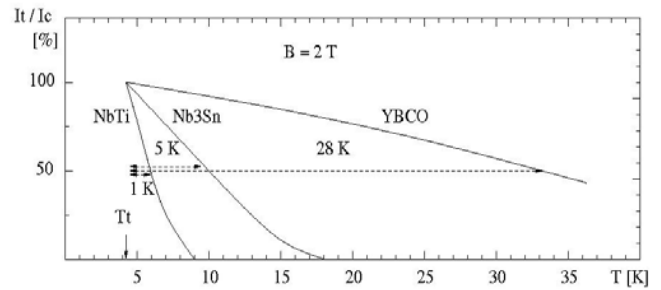

Fig. 1. Temperature margins at various $I_t/I_c$ for NbTi, $Nb_3Sn$ and YBCO superconductors at 2 T field, dashed line is for $I_t/I_c = 50\%$

The LTS wire strands are composed of the thousands of filaments which are embedded in a common copper matrix. This arrangement increases stability against the temperature rise and facilitates the removal of heat. At the cryogenic temperatures, however, the copper's specific heat is low (e.g. ~ 0.1 J/kg-K at 5 K), so the copper matrix absorbs the heat efficiently only for very small temperature rises. The liquid helium, on the other hand, has a very high specific heat (~ 20,000 J/kg-K at 5 K), but as it is not in a direct contact with the filaments, it cools them only via the heat conductivity in the copper matrix. As the copper matrix is typically of (0.6 - 0.8) mm in diameter the cooling response to the filaments in the strand is not only slow but non-uniform, with a preference to the filaments in the most outer layers of the strand.

Contrary to the wire strand, the 344C-2G (YBCO) strand [3] contains only a single filament. This strand is 4.2 mm wide and 200 μm thick. If a space between the strands is allowed for the liquid helium to flow both wide surfaces of the HTS tape will be efficiently cooled. A simplified view of the 344C-2G YBCO strand is shown in Fig.2. The heat path from the

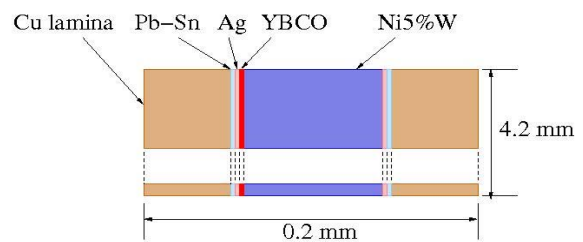

Fig. 2. A simplified view of 344C-2G HTS tape structure

filament to the helium coolant, however, is not symmetric. On one side of the tape it is through a 50 μm thick Cu lamina. On the other side, in addition to the Cu lamina, it must pass

Manuscript received 12 September 2011. This work was supported by Fermi Research Alliance, LLC under DOE Contract DE-AC02-07CH11359.

H. Piekarz, J. Blowers, S. Hays, T. Nicol and V. Shiltsev are with Fermi National Accelerator Laboratory, Batavia, Il 60510, USA (phone: 630-840-2105; fax: 630-840-6039; e-mail: hpiekarz@fnal.gov ).

through the 80 μm thick Ni5%W substrate. Nevertheless, the heat paths from the YBCO filaments to the helium coolant are on average about 3-5 times shorter than those possible with the wire-type strands. In summary, the possibility of having a preferred strand orientation toward the magnetic field, very wide operational temperature margin and a good access of the helium coolant to the individual strand/filament make the HTS tape type superconductors more preferable than the LTS wires for the construction of a fast-cycling magnet power cable.

## II. ARRANGEMENT OF HTS TEST CABLE

A simplified view of the HTS cable arrangement is shown in Fig. 3. The cable consists of 20 344C-2G HTS tapes. The HTS tapes are arranged in 10 pairs with Ni5%W sides facing each other (within the pair) and between each HTS pair a strip of a 4 mm wide, 80 μm thick Ni5%W magnetic screen tape is placed. As discussed in [1], this arrangement may help to suppress the self-field coupling. A space is created between all the HTS strands and the Ni5%W tapes using a 60 μm thick and 6.3 mm wide Kapton tape. These Kapton tape strips which are firmly attached to the surface of both the HTS strand and Ni5%W tape are staggered in the longitudinal direction by 6.3 mm allowing in this way to directly flush half of the strand surface with liquid helium coolant. The HTS stack is firmly squeezed between the two G-10 plates with half-moon cross-sections and inserted into a slit 316LN cryostat pipe of 12.5 mm OD and 0.5 mm wall. The slit pipe is then welded under high tension immobilizing in this way the HTS stack. During the welding a removable aluminum strip under the pipe's slit together with the nitrogen gas purge were used to insure that the temperature of the HTS strands did not exceed $130^0$ C.

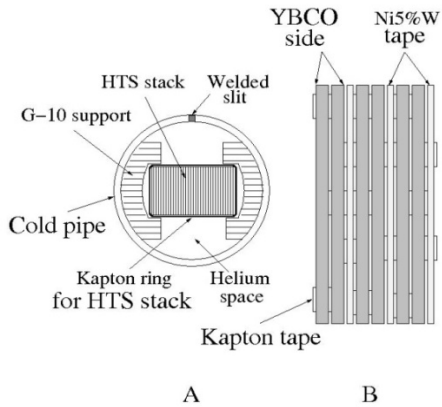

Fig. 3. HTS cable arrangement

## III. ARRANGEMENT OF LTS TEST CABLE

A simplified view of the LTS cable arrangement is shown in Fig. 4. The 27 twisted pairs (5 mm twist pitch) of 0.8 mm diameter NbTi strand (SSC outer dipole) are placed on the outside surface of the densely perforated 316LN pipe (0.9 mm diameter and 0.25 mm wall) serving as the liquid helium channel. This assembly is inserted into a slit 316LN cryostat pipe of 12.5 m diameter and 0.5 mm wall. The slit pipe is welded under tension immobilizing in this way the LTS strands. The temperature of the pipe's surface was controlled in order to minimize the heat induced damage to the strands.

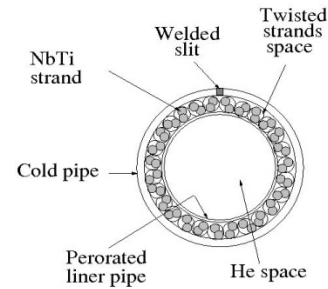

Fig. 4. LTS cable arrangement

The main reason for the arrangement of NbTi strands into the twisted pairs is to minimize their self-field coupling. This arrangement, however, also creates a much larger free space around the strands allowing the liquid helium coolant to flash about a half of each strand's surface.

## IV. TEST ARRANGEMENT AND PROCEDURE

A simplified diagram of the HTS and LTS sub-cables test arrangement is shown in Fig.5. The test cable is placed in the center of the CDA dipole magnet gap. The test cable cryostat can be rotated in X-Y plane within +/- $10^0$. At the inlet and outlet of the test cable helium flow channel there is a pair of Cernox temperature sensors (T1,2 and T3,4) calibrated in (1.4 - 40) K range and Setra pressure transducers (PT1, PT2). The liquid helium from the test cable exits into the 5 m long pipe covered with a 3 kW heater tape. The room temperature helium gas passes then through the flow meter (Brooks, SLA5863) calibrated in the range of (0 – 1000) L/min. Tests were conducted using helium of (8-12) K, 0.23 MPa pressure with the flow rates in [0.4 – 0.8] g/s range. The CDA dipole magnet operated at 0.5 T field with the ramping rates of (2 - 20) Hz producing sweeping magnetic fields of (2 – 20) T/s across the length of the test cable. A typical B-field wave-form produced in the CDA magnet gap is shown in Fig. 6. A characteristic plot of the inlet and outlet cable temperatures, pressure and the flow rate with and without a cycling B-field is shown in Fig.7. The precision of the dissipated power measurement rested mostly on the 0.1 K calibration precision of the temperature sensors thus yielding about 0.25 W for a maximum detectable power loss.

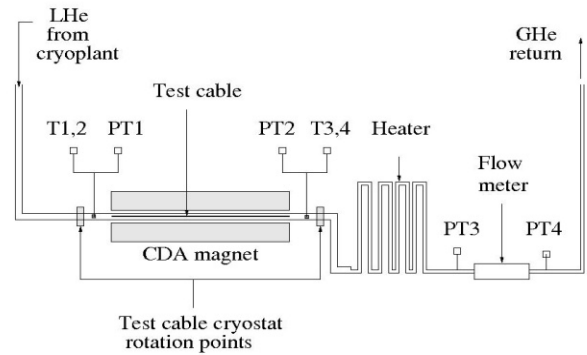

Fig. 5. Test arrangement of the HTS and LTS cables

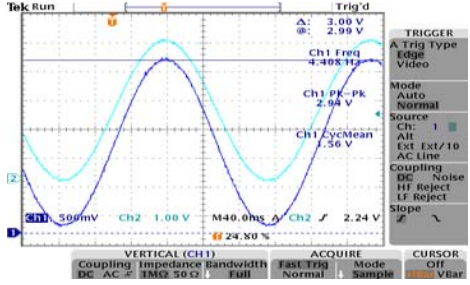

Fig. 6. Wave-forms of CDA magnet current and hall probe in the gap

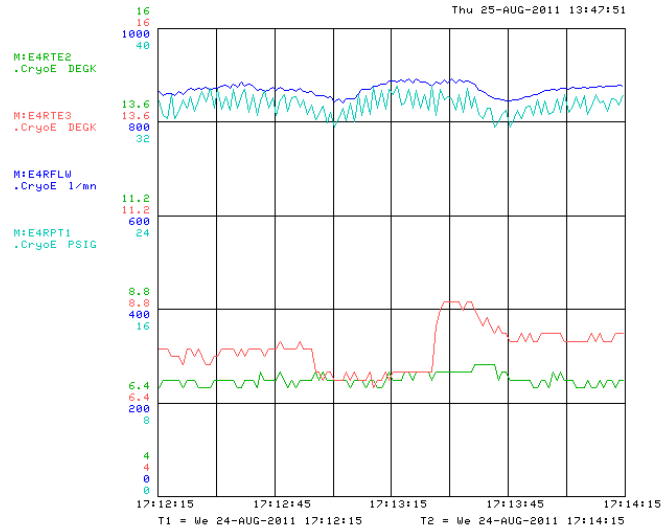

Fig. 7. Typical flow, pressure and out/in temperatures for two different dB/dt rates, part in the centerr is without the sweeping magnetic field.

## V. Projection of Power Losses

The LTS test cable uses the SSC outer dipole NbTi strand (2-F-11010-R17). The VLHC-1 tests [4] have shown that a cable constructed of 276 strands can carry 80 kA DC current at 2 T with 2.5 K operational temperature margin. For the E4R test cable we use 54 strands allowing for about 16 kA DC current with a 2.5 K operational temperature margin. We use the measured power losses of the SIS100 magnet [5] strand (option CSLD) to project losses with the SSC strand. The SIS and SSC strands are wires of 0.8 mm diameter with the Cu/SC = 1.8. They use, however, different filaments: 4 µm and 6 µm diameter, respectively. In addition, the SIS filament twist pitch is 5 mm versus 12.5 mm of the SSC strand. In order to compensate for the SIS smaller twist pitch we twisted the SSC strands (within a pair) using a 5 mm pitch. The SSC strand hysteresis power loss, however, has to be increased relative to the SIS one by a factor of 1.5 (a ratio of 6 µm/4 µm wire diameters).

The estimated critical current for the HTS test cable with 20 344C-2G strands is about 16 KA at 2 T and 15 K [2], thus allowing for about 10 K operational margin. The HTS strand power losses are estimated by scaling the projections for the YBCO filament given in [6], and for the Ni5%W substrate in the strand as well the magnetic screen tapes by using the measurements presented in [7]. The eddy power losses in the LTS and HTS test cables were calculated using the physical parameters of their sub-components as listed in Table 1 and Table 2, respectively. In the HTS strand the B-field induced screening current along the length is $E_Z = -\mu_0 \cdot dB/dt \, (w \cdot \sin\Theta - d \cdot \cos\Theta)$ [8, 9], where $\Theta$ is the angle between the magnetic field and the wide surface of the tape. With the d = 2 µm and w = 4000 µm the ($d \cdot \cos\Theta$) component is negligible, so for the angular dependence of the power losses we use the ($w \cdot \sin\Theta$) term only. The power loss projections are shown as solid lines in Figs. 8, 9.

TABLE I Components of LTS Test-cable

| LTS cable | Unit cross-section [mm$^2$] | Volume/cable [m$^3$] | Mass/cable [g] | Resistivity (5 K) [Ω·m] |
|---|---|---|---|---|
| NbTi | 0.185 | 119.9 10$^{-7}$ | 78.4 | - |
| Balance Cu | 0.328 | 212.6 10$^{-7}$ | 190.4 | 8 10$^{-10}$ |
| Liner pipe | 6.7 | 8.1 10$^{-6}$ | 62.7 | 5 10$^{-7}$ |
| Outer pipe | 18.1 | 21.7 10$^{-6}$ | 169.4 | 5 10$^{-7}$ |
| Cryopipe total | 24.8 | 29.8 10$^{-6}$ | 232.1 | 5 10$^{-7}$ |
| Cable total | - | 92.9 10$^{-6}$ | 500.9 | - |
| He flow area | 77 | | | |

TABLE II Components of HTS Test-cable

| HTS cable | Unit cross-section [mm$^2$] | Volume/cable [m$^3$] | Mass/cable [g] | Resistivity (5K) [Ω·m] |
|---|---|---|---|---|
| YBCO | 0.008 | 19.2 10$^{-8}$ | - | - |
| Ni5%W | 0.320 | 111.4 10$^{-7}$ | 109.8 | 2.55 10$^{-7}$ |
| Ag cap | 0.024 | 57.6 10$^{-8}$ | 6.5 | 1 10$^{-10}$ |
| Cu cap | 0.420 | 96.7 10$^{-7}$ | 86.6 | 8 10$^{-10}$ |
| 344C-2G | 0.772 | 21.6 10$^{-6}$ | 202.9 | - |
| Cryopipe | 18.1 | 29.8 10$^{-6}$ | 169.4 | 5 10$^{-7}$ |
| Cable total | | 73.0 10$^{-6}$ | 372.3 | - |
| He flow area | 67 | | | |

## VI. Test Results

The measured and projected power losses for the LTS and HTS test cables as a function of the dB/dt rate are shown in Fig. 8. In the dB/dt range up to 8 T/s the HTS test cable power losses are about 5 times lower than those with the LTS one. The power losses as a function of the HTS stack-to-B-field angle at a dB/dt = 16 T/s are shown in Fig. 9. The LTS data match projections rather well but the HTS data show higher measured losses than predicted. These losses increase linearly with the dB/dt, as illustrated in Fig.10. This additional power loss, however, seems to be independent of the HTS angular orientation (within the 10$^0$ tested range) as the shape of the angular dependence is well reproduced in the data. The linear increase with the dB/dt rate of the unaccounted HTS cable power loss may suggests their hysteresis origin but it is not supported by the angular dependence of power losses where a constant background underneath the data is very suggestive.

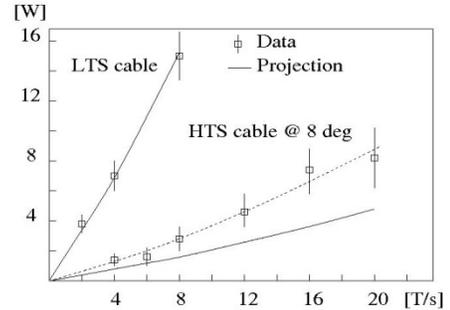

Fig. 8. Measured and projected power losses as a function of dB/dt

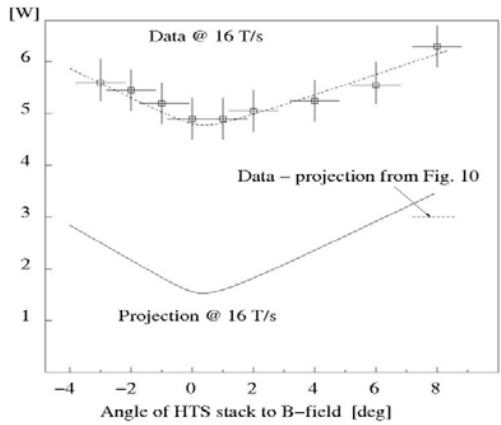

Fig. 9. Power loss at various HTS stack to B-field angles, dashed line indicates data - projection from Fig. 10 at 16 Hz

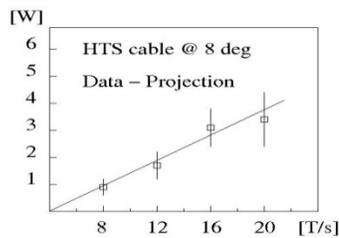

Fig. 10. Power loss data minus prediction for HTS stack at $8^0$

This additional heating-up of the superconducting cable may be caused by the movement of the strands in a cycling magnetic field. In the LTS test cable the strands are very firmly held between the two steel pipes preventing the strand motions. In the HTS stack, however, the 344C tapes are separated from each other with the kapton rings, and perhaps in spite of a very strong transverse compression on the wide surfaces from the G-10 holders some vertical movement of the strands as well as the magnetic screen tapes may occur. The Ni5%W substrates in the HTS strands and magnetic screen tapes are magnetized to a saturation level of about 0.1 T. The magnetic field will then exercise force $F = m \cdot B \cdot \cos\Theta$ on the Ni5%W components, where m is the magnetization strength, B cycling field, and $\Theta$ angle between the field and magnetized component. As the magnetization occurs in the same direction as the field the HTS strands, which are nearly vertically oriented, will be pushed-down and up with each cycle. With the constant value of the magnetic field the force on the strands is constant as well, but the number of mechanical cycles follows the magnetic ones, and so the work and the resulting deposited heat scale with the dB/dt rate. Within the $10^0$ maximum angle range ($\cos 10^0 = 0.985$) and with full magnetization of the Ni5%W strips, one should not expect angular dependence of this kind of power loss.

It was projected in [10] that a superconductor movement by 1/3 µm will generate heat of 750 $J/m^3$. The stored magnetic energy at 0.5 T is $97 \cdot 10^3$ $J/m^3$ thus potentially inflicting a movement of ~43 µm, if the cable was not restrained. The stored magnetic energy in the HTS stack of volume $73 \cdot 10^{-6}$ $m^3$ is 1.1 J, and so at the dB/dt rate of 16 Hz the deposited heat can be up to 17.4 W. The observed unaccounted power loss in the HTS stack at 16 Hz cycle is about 3 W suggesting that the HTS stack is moving by about 7.4 µm ((3W/17.4W)·43 µm). Such a vertical movement of the HTS stack is plausible with the mechanical arrangement used in the test cable.

## VII. SUMMARY AND CONCLUSIONS

The HTS cable suitable for powering an accelerator magnet operating up to 20 T/s has been tested. The power losses of the LTS cable with a comparable allowable transport current were found to be about 5 times higher than those of the HTS ones in the measured range of 8 T/s. The HTS cable power losses, however, are typically about a factor of 2 higher than the projected ones. We believe that the insufficiently restrained movement of the HTS strands (and/or the whole stack) in the vertical plane parallel to the B-field direction is a primary reason for this additional power loss. In the construction of the HTS stack we focused on the immobilization of the transverse movements which would dominate with a transport current present. The effect of the vertical movement can be easily corrected and the new HTS cable design to power the HTS magnet features strong movement restrictions in both vertical and horizontal directions. Consequently, the power losses of fast-cycling magnet with HTS cable can be reduced further, possibly being an order of magnitude lower than with the LTS one, while featuring wide operational temperature margin the same time. The transmission-line type HTS cable can now be considered for application to accelerator magnets with high magnetic field cycling rates.

## ACKNOWLEDGMENTS

We thank Clark Reid and Frank McConolgue for technical designs, and Phil Gallo for the meticulous assembly work. We are grateful to Jerry Makara for operations of E4R cryoplant, and Brad Claypool and Brian Falconet for help with the test system commissioning.